\documentclass[12pt]{article}
\usepackage{graphicx}
\usepackage{amsmath}
\usepackage{verbatim}
\usepackage{subfig}
\usepackage{booktabs}
\newcommand{\mytab}{
\renewcommand{\arraystretch}{0.2}{
\begin{tabular}{lcc}
\toprule
Source & $X(3872) $ & $X(3872)$  \\
 & $K^+\pi^-$ & $K^{0}\pi^{+}$   \\
\midrule
Lepton ID              & 3.4 &  3.4  \\
Kaon ID                & 1.1 &  ...   \\
Pion ID                & 2.5 &  3.2   \\
PDF modeling                 &  $^{+1.8}_{-1.3}$ & $^{+4.2}_{-4.9}$\\
Tracking efficiency          & 2.1 &  2.5 \\
$K^0_S$ reconstruction   &  ...  &  0.7  \\ 
$N_{B\bar{B}}$    & 1.4 & 1.4 \\
Secondary ${\cal B}$   & 0.4 & 0.4  \\
Efficiency & $0.6$ & $1.0$ \\
Fit bias               & 0.6 &  3.1  \\
\midrule
Total & $5.4$  &  $8.0$ \\
\bottomrule
\end{tabular}
}
}

\newcommand{\mytabn}{
\renewcommand{\arraystretch}{0.9}{
\begin{tabular}{lcc}
\toprule
Source & $X(3872) K^*(892)^0$   \\
\midrule
Lepton ID              & 3.4   \\
Kaon ID                & 1.1  \\
Pion ID                & 2.6    \\
PDF modeling                  &  $^{+1.5}_{-1.4}$  \\
Tracking efficiency          & 2.1  \\
$N_{B\bar{B}}$    & 1.4  \\
Secondary ${\cal B}$   & 0.4  \\
MC statistics          & 0.2  \\
Fit bias               &  4.6   \\
\midrule
Total     & $7.0$ \\
\bottomrule
\end{tabular}
}
}

\newcommand\pubdate{\today}

\def\usc{Department of Physics and Astronomy\\
  University of South Carolina, Columbia, SC 29208, USA}
\def\pu{Department of Physics\\
  Panjab University, Chandigarh 160014, India}
\def\kek{Institute of Particle and Nuclear Studies (KEK) \\
  Tsukuba, Ibaraki-ken 305-0801, Japan}
\def\support{\footnote{Speaker on behalf of Belle Collaboration, supported by U.S. Department
of Energy.}}

\def\Title#1{\begin{center} {\Large #1 } \end{center}}
\def\Author#1{\begin{center}{ \sc #1} \end{center}}
\def\Address#1{\begin{center}{ \it #1} \end{center}}

\newenvironment{Abstract}{\begin{quotation}  }{\end{quotation}}
\newenvironment{Presented}{\begin{quotation} \begin{center} 
             PRESENTED AT\end{center}\bigskip 
      \begin{center}\begin{large}}{\end{large}\end{center} \end{quotation}}
\def\Acknowledgements{\bigskip  \bigskip \begin{center} \begin{large}
             \bf ACKNOWLEDGEMENTS \end{large}\end{center}}

\def\beq{\begin{equation}}
\def\eeq#1{\label{#1}\end{equation}}
\def\eeqn{\end{equation}}

\def\beqa{\begin{eqnarray}}
\def\eeqa#1{\label{#1}\end{eqnarray}}
\def\eeqan{\end{eqnarray}}

\let\bar=\overbar

\def\Dslash{\not{\hbox{\kern-4pt $D$}}}
\def\dslash{\not{\hbox{\kern-2pt $\del$}}}

\def\msb{{\bar{\ssstyle M \kern -1pt S}}}

\begin{document}
\begin{titlepage}
\hfill\pubdate

\vfill
\Title{Study of $B \to X(3872) K \pi$ at Belle}
\vfill
\Author{A.Bala$^{a}$, \underline{V. Bhardwaj}\support${^b}$, K. Trabelsi$^c$, J.B. Singh$^a$}
\Address{$^a$\pu \\ $^b$\usc \\ $^c$\kek}
\vfill
\begin{Abstract}
We report the first observation of $B^0 \to X(3872) (K^{+}\pi^{-})$ and
evidence for $B^+ \to X(3872) (K^{0}\pi^{+})$. The product of 
branching fractions for the former decay mode is measured to be ${\cal B}(B^0 \to X(3872) (K^+ \pi^-)) \times {\cal B}(X(3872) \to J/\psi \pi^+ \pi^-) = (7.9 \pm 1.3(\mbox{stat.})\pm 0.4(\mbox{syst.})) \times 10^{-6}$ and also find that
$B^{0}\to X(3872) K^{*}(892)^{0}$ does not dominate the 
$B^{0}\to X(3872)K^{+}\pi^{-}$ decay mode in contrast to other charmonium states 
like $\psi'$. The product of branching fractions for the latter decay mode is measured to be ${\cal B}(B^+ \to X(3872) (K^0 \pi^+)) \times {\cal B}(X(3872) \to J/\psi \pi^+ \pi^-) = (10.6 \pm 3.0(\mbox{stat.}) \pm 0.9(\mbox{syst.})) \times 10^{-6}$. This study is based on the full and final data sample of
711~fb$^{-1}$ ($772\times 10^6 B\bar B$ pairs) collected at the 
$\Upsilon(4S)$ resonance with the Belle detector at the KEKB collider.
\end{Abstract}
\vfill
\begin{Presented}
The 7th International Workshop on Charm Physics (CHARM 2015)\\
Detroit, MI, 18-22 May, 2015
\end{Presented}
\vfill
\end{titlepage}
\def\thefootnote{\fnsymbol{footnote}}
\setcounter{footnote}{0}
%
\section{Introduction}
Belle Collaboration discovered  the $X(3872)$ state~\cite{Choi:2003ue} in 
the exclusive reconstruction of $B^{+} \rightarrow X(3872)(\to J/\psi \pi^+ \pi^-) K^{+}$~\cite{charge_conjugate} about more than a decade ago. Currently, we know precisely its mass 
(3871.69$\pm$0.17)~MeV$/c^2$~\cite{pdg2014}, have 
a stringent limit on its width (less than 1.2~MeV at 90\% confidence 
level)~\cite{Choi:2011prd} along with definitive $J^{PC}$ assignment of
$1^{++}$~\cite{Aaij:2013zoa}. It has been observed to decay to the 
following final states:
$J/\psi \gamma$~\cite{Bhardwaj:2011prl}, 
$\psi' \gamma$~\cite{Aaij:2014pg}, 
$J/\psi \pi^+ \pi^- \pi^0$~\cite{Sanchez:2010prd},
$J/\psi \pi^+\pi^-$~\cite{Choi:2003ue} and
$D^{0} {\bar{D}}^{*0}$~\cite{Aushev:2010prd,Aubert:2008prd_babar}. 
Till now, $X(3872)$ has been observed and studied in two body $B$ meson decays. This is the first time, we have observed $X(3872)$ in three body $B$ decay and estimated the product of branching fractions using full and final Belle data set to understand its mysterious nature. In this analysis, we also did the comparison of this exotic state ``$X(3872)$'' with ordinary charmonium states by considering $\psi'$ as calibration sample.

We present study of $X(3872)$ 
production via the $B^0 \to X(3872) K^+ \pi^-$ and $B^+ \to X(3872) K^0_S \pi^+$ 
decay modes, where the $X(3872)$ decays to 
$J/\psi \pi^+ \pi^-$. The study is based on 711~fb$^{-1}$ of data containing 
$772\times 10^{6}$ $B\bar{B}$ events collected with the Belle 
detector~\cite{Abashian} at the KEKB  
$e^+e^-$ asymmetric-energy collider~\cite{Kurokawa:2003Abe:2103} operating 
at the $\Upsilon(4S)$ resonance. 

\section{Selection criterion}

\par To find the reconstruction efficiencies, signal Monte Carlo (MC) samples are generated for each decay mode using 
EvtGen~\cite{x_evtgen} and radiative effects are taken into account using the PHOTOS~\cite{x_photos} package. The detector response is 
simulated using Geant3~\cite{x_geant}. The selection criteria is same for signal MC events, background MC events and data events for calibration sample (having $\psi'$) and for concerned decay modes (having $X(3872)$) except the difference of $M_{J/\psi\pi^+\pi^-}$ range as both ($\psi'$ and $X(3872)$) are further reconstructed from $J/\psi\pi^+\pi^-$.
\par We reconstruct $J/\psi$ mesons in the $\ell^+ \ell^-$ decay channel
($\ell =e~\rm {or}~\mu$) and include bremsstrahlung 
photons that are within 50 mrad of either the $e^+$ or $e^-$ tracks 
[hereinafter denoted as $e^+ e^- (\gamma)$].
The invariant mass of the $J/\psi$ candidate is required 
to satisfy $3.00$~GeV/$c^2$ $< M_{e^+ e^- (\gamma)} <  3.13$~GeV/$c^2$ 
or $3.06$~GeV/$c^2 < M_{\mu^+ \mu^-}  <  3.13$~GeV/$c^2$ (with a distinct lower 
value accounting for the residual bremsstrahlung in the electron mode). 
The $J/\psi$ candidate is then combined with a $\pi^+\pi^-$ pair to 
form an $X(3872)$ ($\psi'$) candidate whose mass must satisfy 
3.82~GeV$/c^2$ $<$ $M_{J/\psi\pi^+\pi^-}$ $<$ 3.92~GeV$/c^2$
(3.64~GeV$/c^2$ $<$ $M_{J/\psi\pi^+\pi^-}$ $<$ 3.74~GeV$/c^2$).
The dipion mass must also satisfy 
$M_{\pi^+\pi^-} > M_{J/\psi\pi^+\pi^-} - (m_{J/\psi} + 0.2~{\rm GeV}/c^2)$, where 
$m_{J/\psi}$ is nominal mass. This criterion corresponds to 
$M_{\pi^+\pi^-} > 575~(389)$~MeV/$c^2$ for the $X(3872)~(\psi')$ mass region and
it reduces significantly the combinatorial background~\cite{Choi:2011prd}
with an advantage of flattening the background distribution
in $M_{J/\psi \pi^+ \pi^-}$. To suppress the background from
continuum events, we require $R_2 < 0.4$, 
where $R_2$ is the ratio of the second- to zeroth-order Fox-Wolfram 
moments~\cite{foxwolf}.  

\par To reconstruct neutral (charged) $B$ meson candidate, a $K^+\pi^-$ 
($K^0_S \pi^+$) candidate is further combined with the $X(3872)$ for concerned decay mode and with $\psi'$ for the study of calibration sample. $B$ candidates are selected using two kinematic variables: the 
energy difference $\Delta E = E_B^* - E_{\rm beam}$ and the beam-energy 
constrained mass $M_{\rm bc} = (\sqrt{E_{\rm beam}^{2}- p_B^{*2}c^2})/c^2$, where 
${E_{ \rm beam}}$ is the beam energy and $E_B^*$ and $p_B^*$ are 
the energy and magnitude of momentum, respectively, of the candidate 
$B$-meson, all calculated in the $e^+ e^-$ center-of-mass (CM) frame.
More details regarding the selection criteria
can be found in Ref.~\cite{AnuBala}.
\section{Signal Extracton}
To extract the signal yield of $B\to X(3872)(\to J/\psi\pi^+\pi^-) K\pi$,
 we perform a two-dimensional (2D) unbinned extended maximum 
likelihood fit to the $\Delta E$ and $M_{J/\psi\pi\pi}$ distributions.
The 2D probability distribution function (PDF) is a product of the individual 
one-dimensional PDFs, as no sizable correlation is found.
\par In order to study backgrounds,
 we use a large Monte Carlo sample of 
$B \to J/\psi X$ events, corresponds to 100 times the integrated luminosity 
of the data sample. Based upon above study we find that few backgrounds are peaking in the $M_{J/\psi\pi\pi}$ 
distribution (nonpeaking in the $\Delta E$ distribution) and vice versa. The remaining backgrounds are 
combinatorial in nature and are flat in both distributions. 
\par
\begin{figure}[!ht]
\begin{center}
  \begin{tabular}{cc} 
\includegraphics[height=38mm,width=120mm]{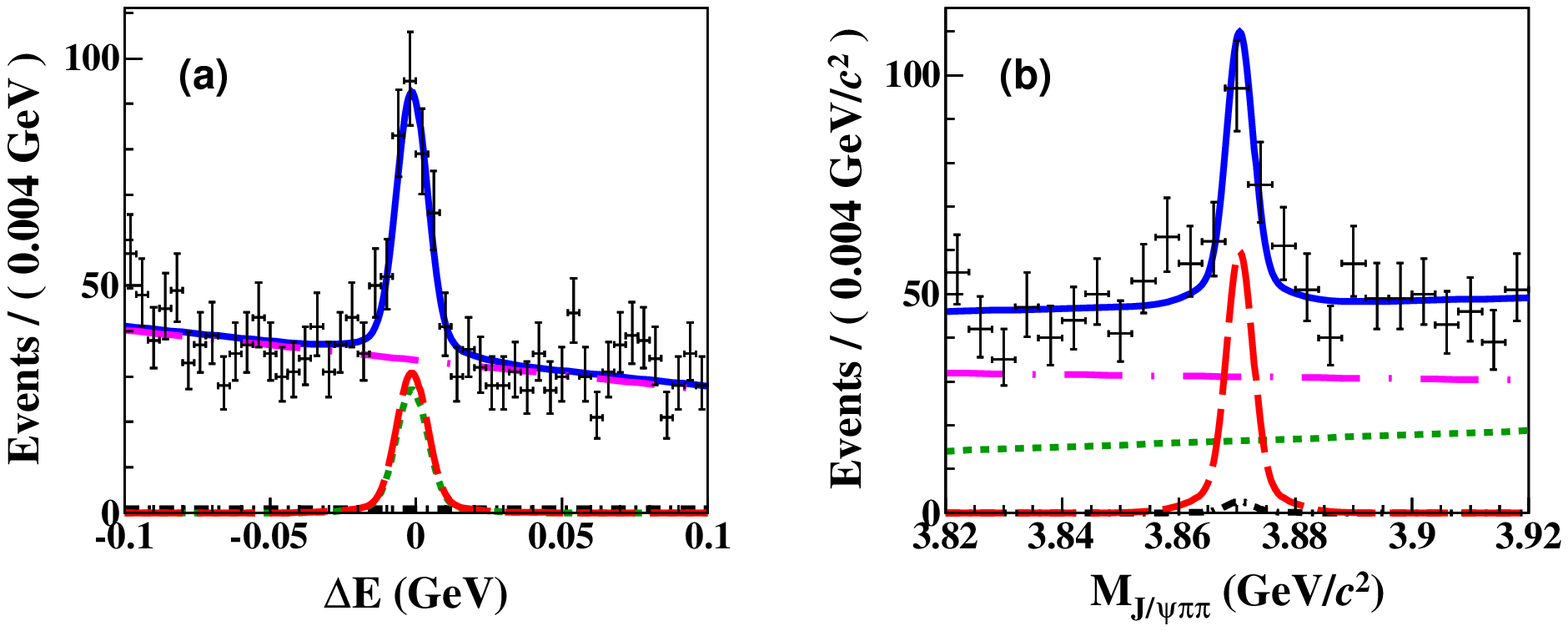}\\
\includegraphics[height=38mm,width=120mm]{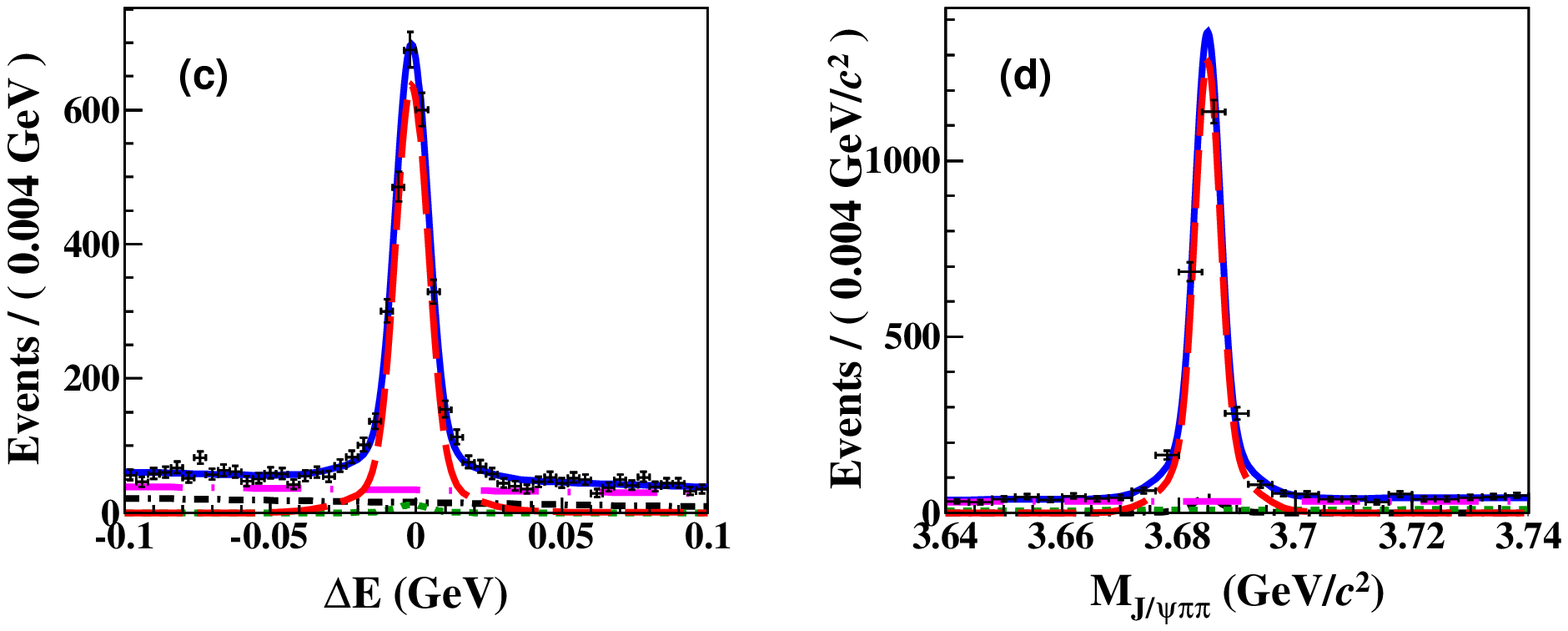}
   \end{tabular}
\caption{Projections of 
(a) $\Delta E$ distribution with 
3.860~GeV$/c^2< M_{J/\psi\pi\pi} <$ 3.881~GeV$/c^2$, 
(b) $M_{J/\psi\pi\pi}$ distribution with $-11$~MeV $< \Delta E <$ 8~MeV,
(c) $\Delta E$ distribution with 3.675~GeV$/c^2< M_{J/\psi\pi\pi} <$ 
3.695~GeV$/c^2$, and (d) $M_{J/\psi\pi\pi}$ distribution with 
$-11$~MeV $< \Delta E <$ 8~MeV  for
$B^0\to X(3872)(\to J/\psi\pi^+\pi^-)K^+\pi^-$ decay mode (top) and the 
$B^0\to \psi'(\to J/\psi\pi^+\pi^-)K^+\pi^-$ decay mode (bottom).
The curves show the signal [red long-dashed]
  and the background components [
black dashed-dot for the component peaking in $M_{J/\psi\pi\pi}$ 
but non-peaking in $\Delta E$, green dashed for the one peaking in $\Delta E$ 
but non-peaking in $M_{J/\psi\pi\pi}$, and magenta long dashed-dot 
for combinatorial background] as well as the overall fit [blue solid].}
\label{fig:signal_enhanced_psi}
\end{center}
\end{figure}

 
For the signal, the $\Delta E$ dimension parametrization is done by the sum of a Crystal 
Ball~\cite{crystal_ball} and a Gaussian function while the $M_{J/\psi\pi\pi}$ 
distribution is modeled using the sum of two Gaussians having a common mean.  
 The mean and resolution of $\Delta E$ and $M_{J/\psi\pi\pi}$ are fixed 
for the $X(3872)$ mass region from signal MC samples after being rescaled from the 
results of the $B^{0}\to\psi'K^{+}\pi^{-}$ decay mode. Further, we correct 
the mean of a Gaussian function for the $M_{J/\psi\pi\pi}$ distribution because of difference between the decay dynamics of $\psi'$ and $X(3872)$. 
The tail parameters are fixed according to the signal MC simulation. The 
peaking components can be divided into two categories: the one peaking in 
$M_{J/\psi\pi\pi}$ but non-peaking in $\Delta E$ 
that comes from the $B\to X(3872) X'$ decays where the $X(3872)$ decays in 
$J/\psi\pi^+\pi^-$ [here $X'$ can be any particle], and the other peaking in $\Delta E$ but non-peaking 
in $M_{J/\psi \pi \pi}$ which comes from a $B$ with the same final state 
where $J/\psi\pi^+\pi^-$ is not from a $X(3872)$. 
The peaking background in $\Delta E$ ($M_{J/\psi \pi \pi}$) is found to have the
same resolution as that of the signal, so the PDF is chosen to be the same 
as the signal PDF, while the non-peaking background in the other dimension 
is parameterized with a first-order Chebyshev polynomial. For the 
combinatorial background in both dimensions, a first-order Chebyshev
polynomial is used.
The fits are first validated on full simulated experiments and toy MC studies and no significant bias 
is seen. Fig.~\ref{fig:signal_enhanced_psi}~(top)
shows the signal-enhanced projection plots for the $B^0\to X(3872) (K^+ \pi^-)$ decay mode. 
The result of the fit and branching fractions derived are listed in 
Table~\ref{table:tab_results}.
We observe a clear signal for $B^{0}\to X(3872)K^{+}\pi^{-}$  with $116 \pm 19$ signal events corresponding to a significance (including systematic uncertainties related to the signal yield as mentioned in Table~\ref{table:tab_results}) of 7.0 standard deviations 
($\sigma$), and measure the product of branching fractions to be
$\mathcal{B}(B^{0}\to X(3872)K^{+}\pi^{-}) \times \mathcal{B}(X(3872)\to J/\psi\pi^{+}\pi^{-}) = (7.9 \pm 1.3 (\mbox{stat.}) \pm 0.4 (\mbox{syst.}))\times 10^{-6}$.
The above fit is validated on the calibration mode $B^0\to\psi'K^+\pi^-$. 
Fig.~\ref{fig:signal_enhanced_psi} (bottom) shows the signal-enhanced 
projection plots for the $B^0\to \psi' (K^+ \pi^-)$ decay mode. 
We measure the branching fraction to be 
$\mathcal{B}(B^{0} \to \psi'K^{+}\pi^{-})$ =
$(5.79 \pm 0.14(\mbox{stat.}))\times 10^{-4}$,
consistent with an independent Belle result based on a Dalitz-plot 
analysis~\cite{Chilikin:2013tch}.

\begin{table*}[!hbpt]
\caption{Signal yield (Y) from the fit, weighted efficiency ($\epsilon$) after PID 
correction, significance ($\Sigma$) and measured $\mathcal{B}$ for 
$B^0\to X(3872)K^+\pi^-$ and $B^+\to X(3872)K^0\pi^+$.
The first (second) uncertainty represents a statistical 
(systematic) contribution.}
\begin{center}
\renewcommand{\arraystretch}{0.2}{
\begin{tabular}{lcccc}
\hline
\hline
Decay Mode &  Yield (Y)$\;$ & $\;$ $\epsilon$ (\%) &
$\;$ $\Sigma$ ($\sigma$) &
 ${\cal B}{(B \to X(3872) K \pi)} \times$ \\
 &  &  &  & ${\cal B}(X(3872) \to J/\psi \pi^+ \pi^-)$ \\
\hline
$B^0 \to X(3872) K^+\pi^-$ & $116 \pm 19$ & 15.99 & 7.0 & $(7.9 \pm 1.3 \pm 0.4)\times 10^{-6}$ \\
$B^+ \to X(3872) K^{0}\pi^{+}$ & $35 \pm 10$ & 10.31 & 3.7 & $(10.6 \pm 3.0 \pm 0.9) \times 10^{-6}$ \\
\hline
\hline
\end{tabular}
}
\end{center}
\label{table:tab_results}
\end{table*}
\par
\begin{figure}[!htbp]
\begin{center}
  \begin{tabular}{cc}  
\includegraphics[height=55mm,width=70mm]{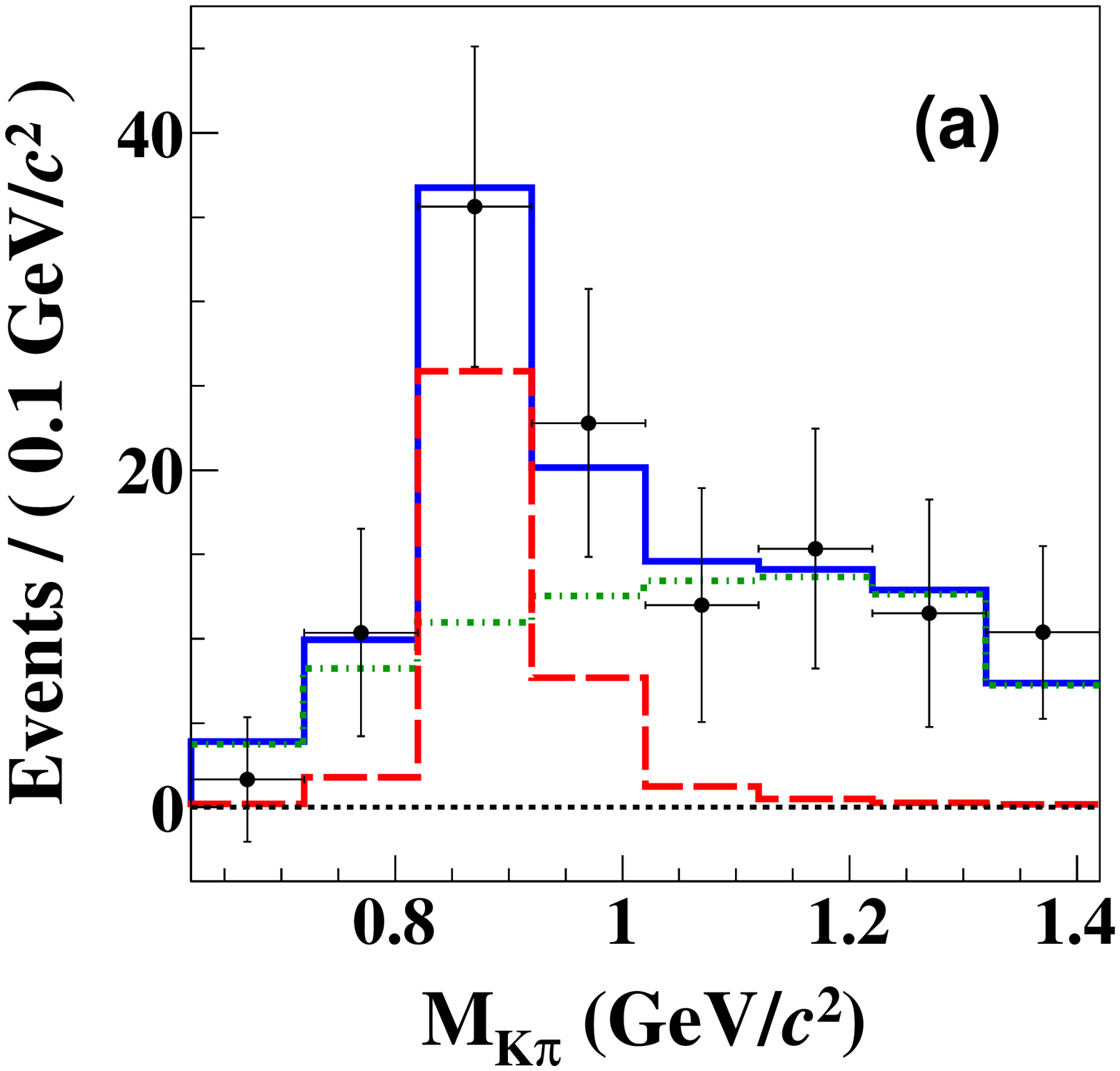}
\includegraphics[height=55mm,width=70mm]{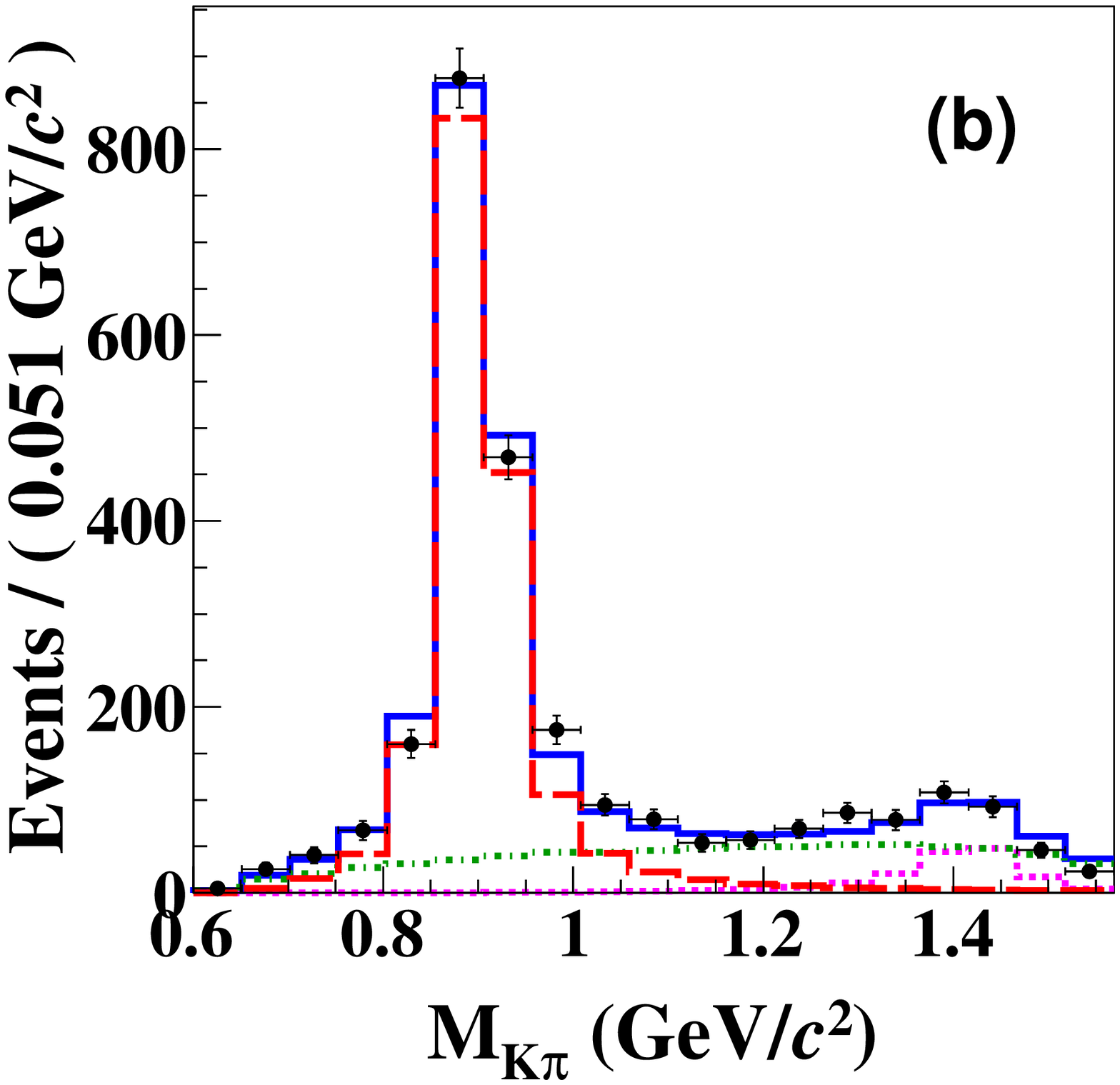}
  \end{tabular}
  \caption{Fit to the background-subtracted $M_{K\pi}$ distribution: 
    (a) for the $B^0\to X(3872) (K^+\pi^-)$ decay mode, the curves show the
    $B^0\to X(3872) K^*(892)^0$ [red long-dashed], 
    $B^0\to X(3872) (K^+\pi^-)_{\rm NR}$ [green dot-dashed],
    as well as the overall fit [blue solid].
    (b) for the $B^0\to \psi' (K^+\pi^-)$ decay mode, the curves show the 
    $B^0\to \psi' K^*(892)^0$ [red long-dashed], 
    $B^0\to \psi' (K^+\pi^-)_{\rm NR}$ [green dot-dashed],
    $B^0\to \psi' K_{2}^*(1430)^{0}$ [magenta dashed]
    as well as the overall fit [blue solid], 
    }
  
\label{fig:data_binned_psi}
\end{center}
\end{figure}

 Further, to determine the contribution of the $K^*(892)$ and other intermediate states,
we perform a 2D fit to $\Delta E$ and $M_{J/\psi\pi\pi}$ in each bin of $M_{K\pi}$ 
(100-MeV wide bins of $M_{K\pi}$ in the range $[0.62, 1.42]$ GeV/$c^2$) for $X(3872)$ mass region, 
which provides a background-subtracted $M_{K \pi}$ signal distribution.
All parameters of the signal PDFs for $M_{J/\psi\pi\pi}$ and $\Delta E$ 
distributions are fixed from the previous 2D fit to all events. 
Then we perform a binned minimum $\chi^2$ fit to  the $M_{K\pi}$  distribution using 
$K^*(892)^0$ and $(K^{+}\pi^{-})_{\rm NR}$ components, which are histogram PDFs
obtained from MC samples. Note that the 
$B^{0}\to X(3872) {K_{2}}^{*}(1430)^{0}$ decay is kinematically suppressed. 
The resulting fit result is shown in Fig.~\ref{fig:data_binned_psi}(a).
We obtain $38 \pm 14$ ($82 \pm 21$) signal events for the
$B^{0}\to X(3872) K^{*}(892)^{0}$ ($B^{0}\to X(3872) (K^+\pi^-)_{\rm NR}$) decay 
mode. 
This corresponds to a 3.0$\sigma$ significance (including 
systematic uncertainties related to the signal yield) for the
$B^{0}\to X(3872)(\to J/\psi \pi^{+} \pi^{-}) K^{*}(892)^0$ decay mode, 
and a product of branching fractions of 
${\cal B}(B^{0} \to X(3872)K^*(892)^0) \times {\cal B}(X(3872) \to J/\psi \pi^+ \pi^-)  = (4.0 \pm 1.5(\mbox{stat.}) \pm 0.3(\mbox{syst.})) \times 10^{-6}$.
The ratio of branching fractions is:
\begin{equation}
\label{eqn:fraction}
\begin{split}
\frac{\mathcal{B}(B^0 \to X(3872)K^{*}(892)^{0})\times \mathcal{B}(K^{*}(892)^{0} \to K^{+}\pi^{-})}{\mathcal{B}(B^0 \to X(3872) K^+\pi^- )} \\
 = 0.34 \pm 0.09(\mbox{stat.}) \pm 0.02(\mbox{syst.}).
\end{split}
\end{equation}
In the above ratio, all systematic uncertainties cancel except 
those from the PDF model, fit bias and efficiency variation over the Dalitz distribution.
\par The same procedure is also applied to the $B^0 \to \psi' K^+ \pi^-$ mode. 
With the sufficient yield, we use 51-MeV wide bins of $M_{K\pi}$ 
in the range [0.600, 1.569]~GeV$/c^2$.
We perform a binned minimum $\chi^2$ fit to the obtained $M_{K\pi}$ 
signal distribution again to extract the contributions of the $K\pi$ 
non-resonant and resonant components. For this purpose, we use 
histogram PDFs obtained from MC samples of several possible components of the $(K^+\pi^-)$ system: ${K}^{*}(892)^{0}$, 
$K_2^*(1430)^{0}$ and non-resonant $K^+\pi^-$ $((K^{+}\pi^{-})_{\rm NR})$.
The fit result is shown in Fig.~\ref{fig:data_binned_psi}(b).
The $K^*(892)^0$ component dominates and we measure 
$\mathcal {B}(B^{0}\to\psi'{K}^{*}(892)^{0}) = 
(5.88 \pm 0.18 (\mbox{stat.})) \times 10^{-4}$,  
which is consistent with the world average~\cite{pdg2014}. 
\par In contrast to $B^{0}\to \psi' (K^+\pi^-)$ (where the ratio 
of branching fractions is $0.68 \pm 0.01(\mbox{stat.})$), 
$B^{0}\to X(3872) K^{*}(892)^{0}$ is not dominating in
the $B^{0}\to X(3872)K^{+}\pi^{-}$ decay mode. 
\begin{figure}[!h]
\centering
  \begin{tabular}{cc}  
\includegraphics[height=37mm,width=125mm]{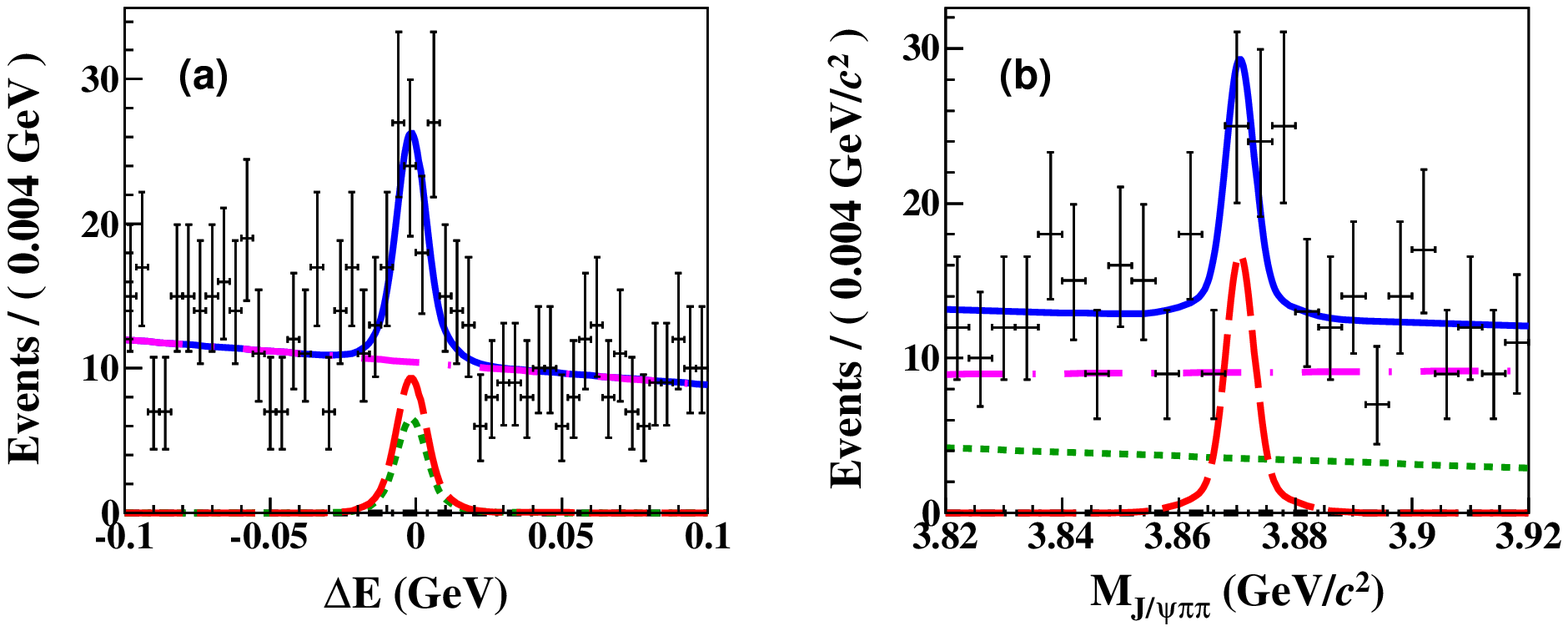}\\
\includegraphics[height=37mm,width=125mm]{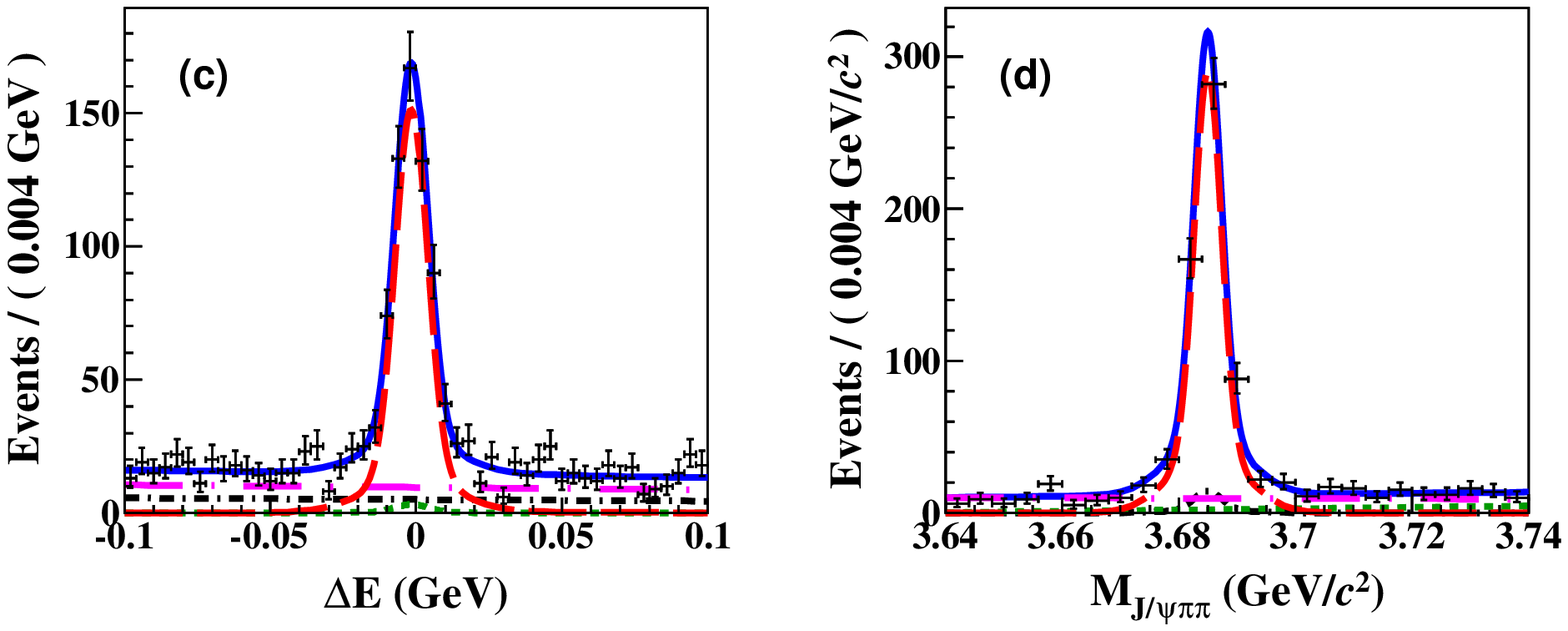} 
     \end{tabular}
  \caption{Projections of 
    (a) $\Delta E$ distribution with 3.859~GeV$/c^2< M_{J/\psi\pi\pi} <$ 
    3.882~GeV$/c^2$ and
    (b) $M_{J/\psi\pi\pi}$ distribution with $-11$~MeV $< \Delta E <$ 9~MeV,
    (c) $\Delta E$ distribution with 3.675 GeV$/c^2< M_{J/\psi\pi\pi} <$ 
    3.695~GeV$/c^2$,
    (d) $M_{J/\psi\pi\pi}$ distribution with $-11$~MeV $< \Delta E <$ 9~MeV
    for the $B^\pm \to X(3872)(\to J/\psi\pi^+\pi^-)K_S^{0}\pi^{\pm}$ decay mode 
    (top) 
    and for the $B^{\pm}\to \psi'(\to J/\psi\pi^{+}\pi^{-})K_S^{0}\pi^{\pm}$ decay 
    mode (bottom). Color representation is same as that of neutral mode.}
  \label{fig:2d_data_ks}
\end{figure}

\par We also investigate the decays 
$B^{+}\to X(3872)(\to J/\psi \pi^{+} \pi^{-})(K^{0}\pi^{+})$.
We perform a 2D fit to $\Delta E$ and $M_{J/\psi\pi\pi}$, as before. 
The projections of the 2D fit for $B^{+}\to X(3872)(\to J/\psi \pi^{+} \pi^{-}) (K^{0}\pi^{+})$ in the
signal-enhanced regions are shown in Figs.~\ref{fig:2d_data_ks}(a) and (b). 
We find $35 \pm 10$ events for the
$B^{+}\to X(3872)(\to J/\psi \pi^{+} \pi^{-}) (K^{0}\pi^{+})$ decay mode, 
corresponding to a 3.7$\sigma$ significance (including 
systematic uncertainties). The product of branching fractions is 
${\cal B}(B^{+} \to X(3872)K^0\pi^{+}) \times {\cal B}(X(3872) \to J/\psi \pi^+ \pi^-)  = (10.6 \pm 3.0(\mbox{stat.}) \pm 0.9(\mbox{syst.})) \times 10^{-6}$.
The above fit is validated for the $\psi'$ mass region.
The projections of the 2D fit for $B^{+}\to \psi'(\to J/\psi \pi^{+} \pi^{-}) (K^{0}\pi^{+})$ in the
signal-enhanced regions are shown in Figs.~\ref{fig:2d_data_ks}(c) and (d). 
The branching fraction for 
$B^{+}\to \psi'(\to J/\psi \pi^{+} \pi^{-}) (K^{0}\pi^{+})$ 
is $(6.00 \pm 0.28 (\mbox{stat.})) \times 10^{-4}$, while the 
world average of this quantity is $(5.88 \pm 0.34)\times 10^{-4}$. 
\par Systematic uncertainties are summarized in Table~\ref{tbl:table}. All systematic uncertainties are added in quadrature to give total systematic uncertainty of 5.4\%, 8.0\%, 7.0\% for $B^0\to X(3872) K^+ \pi^-$, $B^+\to X(3872) K_S^0 \pi^+$ 
and $B^0\to X(3872) K^*(892)^0$, respectively.
\begin{table}[!h]%
  \subfloat[][]{\mytab}%
  \qquad
  \subfloat[][]{\mytabn}
  \caption{Summary of the systematic uncertainties in percent (a) used for 2D fit. (b) used for the 
$M_{K\pi}$ background-subtracted fit in $B^0\to X(3872) K^+ \pi^-$.}%
  \label{tbl:table}%
\end{table}
\par In summary, we report the first observation of the $X(3872)$ in the decay 
$B^0 \to X(3872)K^+\pi^-$, $X(3872) \to J/\psi \pi^+ \pi^-$. 
The result for the $X(3872)$, where $B^0 \to X(3872) K^*(892)^0$
does not dominate the  $B^0 \to X(3872) (K^+\pi^-)$ decay, is in marked
contrast to the $\psi'$ case.
We have checked for a structure in 
the $X(3872)\pi$ and $X(3872)K$ invariant masses and found no evident peaks.
We measure ${\cal B}(B^0 \to X(3872) (K^+ \pi^-)) 
\times {\cal B}(X(3872) \to J/\psi \pi^+ \pi^-) = 
(7.9 \pm 1.3(\mbox{stat.}) \pm 0.4(\mbox{syst.}))\times 10^{-6}$
and ${\cal B}(B^+ \to X(3872)K^{0}\pi^{+}) 
\times {\cal B}(X(3872) \to J/\psi \pi^+ \pi^-) = (10.6 \pm 3.0(\mbox{stat.}) \pm 0.9(\mbox{syst.})) \times 10^{-6}$. \\


\begin{thebibliography}{99}
\bibitem{charge_conjugate}
Charge-conjugate decays are included unless explicitly
stated otherwise.
\end{thebibliography}
\end{document}